\documentclass[pre,preprint,showpacs]{revtex4}
\usepackage{epsfig}
\usepackage{amssymb}
\usepackage{amsfonts}
\usepackage{amsthm}
\usepackage{newlfont}
\usepackage{epsfig}
\usepackage{amscd}
\usepackage{graphics}

\newcommand{\beq}{\begin{equation}}
\newcommand{\eeq}{\end{equation}}
\newcommand{\ba}{\begin{array}}
\newcommand{\ea}{\end{array}}
\newcommand{\bea}{\begin{eqnarray}}
\newcommand{\eea}{\end{eqnarray}}
\begin{document}

\title{The high probability
 state transfers and entanglements between different nodes of the  homogeneous  spin 1/2 chain in 
inhomogeneous external magnetic field}
\author{S.I.Doronin and A.I.Zenchuk}
\email{ zenchuk@itp.ac.ru}
\affiliation{
Institute of Problems of Chemical Physics, Russian Academy of Sciences, Chernogolovka, Moscow reg., 142432, Russia}

\date{\today}

\begin{abstract}
We consider the  high probability state transfers and entanglements between different nodes of the spin-1/2 chains governed by the XXZ-Hamiltonian using the inhomogeneous stationary  external magnetic field. Examples of three-, four-, ten- and twenty-node chains are represented. A variant of  realization of the proper inhomogeneous magnetic field is discussed.
\end{abstract}

\pacs{
}

\maketitle

\section{Introduction}
The problem of either perfect state transfer (PST) \cite{Bose,CDEL,ACDE,KF,KS} or high probability state transfer (HPST) \cite{KZ,GKMT} is closely related to the construction of   quantum communication channels. Essentially, the effect of quantum state transfer in long homogeneous  chains (of 1000 and 10000 nodes) has been observed in numerical experiments with so-called "quantum echo"   \cite{FBE}, although the probability of those transfers is not high.  At present time, most efforts are aimed on   study of the end-to-end quantum state transfer. In particular, it is shown  that  symmetrical inhomogeneous chains are most suitable for this purpose \cite{CDEL,ACDE,KS,KF,GMT,GKMT}. There are several different methods allowing one to perform the  end-to-end arbitrary quantum state transfer in the long spin chains. We list those methods which are well known and relevant to our paper:
\begin{enumerate}
\item The  homogeneous chains with XXZ Hamiltonian having week end bonds \cite{GKMT} provide the HPST. The dipole-dipole interactions among all nodes have been taken into account.
\item The symmetrical chains with XY Hamiltonian having coupling constants increasing properly from the ends to the center of the chain \cite{CDEL,ACDE,KS}. It is shown that such chains provide the PST if only the approximation of nearest neighbor interaction is used. However, attempt to organize the PST using all interactions in inhomogeneous chains causes serious  problem \cite{Kay}. 
\item The chains with XY Hamiltonian and alternating coupling constants \cite{KF,KZ} provide the HPST except the four-node chain where the PST is possible (the approximation of nearest neighbor interaction has been used).
\item
The information-flux approach to the quantum state transfer allows one to arrange the PST using the inhomogeneous spin chains with either Ising or XY Hamiltonians and the approximation of nearest neighbor
interaction   \cite{FPPK,FPK}.
\item
The homogeneous chain with  switching  magnetic field at the receiving end node of the chain governed by either Heisenberg or XY Hamiltonian \cite{BGB} allows one to organize the HPST using the approximation of nearest neighbor
interaction.
\end{enumerate}
The problem of end-to-end entanglement in both homogeneous and inhomogeneous  one-dimensional chains has also been studied up to some extent  \cite{Bose,VGIZ,GMT}.

Nevertheless, the practical implementation of the quantum state transfer in the above chains is difficult  because one has  either to construct chains with given set of coupling constants (see the 1st-4th items above) or organize the switching  magnetic field at the receiving end node (see the 5th item above). 
The problem of the quantum state transfer to inner nodes of the chain  is even more complicated. However, this problem is important both for the construction of quantum registers and for the information distribution among different 
receivers. 
Two different methods allowing one to  distribute  the quantum state among many different nodes of spin-1/2 system  are suggested in \cite{FZ} (the line chains of $2^n$ nodes with  week bonds which are properly distributed along the chain; here $n$ is an arbitrary positive integer) and \cite{DFZ} (the simplest two- and three-dimensional spin systems of four and eight nodes respectively). 

 Other  problems requiring further study are the realization of the strong entanglement between any two nodes of the quantum system and the relation between entanglements and quantum state transfer probabilities.
In particular, it is shown  that the HPST between the $n$th and the $m$th nodes does not always require the strong entanglement between these nodes \cite{GMT,DFZ}.

Usually, quantum state transfers and entanglements are studied   in the strong homogeneous magnetic field. However, inhomogeneous magnetic field may be also helpful. Thus, a Heisenberg model with the parabolic magnetic field has been used for the wave packet transfer along the chain (rather than for the single state transfer) \cite{SLSS}. 

In this paper, we suggest an alternative method providing both the arbitrary quantum state transfer  and the entanglement between different nodes (not only between the end nodes)  of  one-dimensional homogeneous spin chains
using a stationary inhomogeneous magnetic field. 
We  show that, in general, the HPST between the $n$th and the $m$th nodes  causes  the strong entanglement between these nodes,
however the time moment corresponding to the strong entanglement may be very large. The opposite is also true, i.e.
the strong  entanglement
 between the $n$th and the  $m$th nodes  at some  time moment 
causes the HPST but the time interval required for such transfer may be very long.

 We will use model with dipole-dipole interactions among all nodes throughout this paper. This is important, because approximation of nearest neighbor interactions
 yields wrong results in the case of dipole-dipole interactions. In particular,  the state transfer time intervals become
 much longer in comparison with those which have been obtained using all interactions.

 Although we do not give a proper theoretical explanation of the mechanism of such transfers (some explanation is given in Sec.\ref{Section:Hamiltonian}, see 
eq.(\ref{perfect})), we represent results of the numerical simulations for the chains of three, four, ten and twenty nodes.
Remark, that the problem of state transfer (and entanglement) between any nodes is nontrivial even for short chains. In particular, it is impossible to realize the HPST from the first to the second node of the three node chain in homogeneous magnetic field  plying with coupling constants only. For this reason, we consider the state transfer problem   for the short chains of three and four nodes. Regarding longer chains (of ten and twenty nodes) 
we represent  only results concerning the end-to-end HPST and the realization of the strong  entanglement between
the  end nodes.  Note, nevertheless, that the state transfers to inner nodes of such chains have also been realized in our numerical simulations.

We also suggest a method for generation of  inhomogeneous magnetic field  using a system of  direct currents properly distributed with respect to the spin chain.

  The basic advantage of our approach to the quantum state transfer problem 
   is that the transmission line is represented by 
 the homogeneous chain which may be produced  simpler then inhomogeneous one. The HPSTs and entanglements  between different nodes  (not only between the end nodes) in such transmission line are  organized by   some additional environment generating the proper inhomogeneous magnetic field.

We recall the definitions of  HPST and  entanglement in the spin-1/2 chain with XXZ Hamiltonian in Sec.\ref{Section:Hamiltonian}. A method allowing one to generate the required distribution of the  inhomogeneous magnetic field is discussed in Sec.\ref{Section:mf}. The HPSTs and entanglements between  different nodes of the three- and  four-node chains as well as between the end nodes of the ten- and  twenty-node chains  are discussed in Sec.\ref{Section:examples}. Finally, we give conclusions in Sec.\ref{Section:conclusions}.

\section{HPSTs and entanglements in the spin-1/2 chains with XXZ Hamiltonian}
 \label{Section:Hamiltonian}
We study the quantum state transfers    among  different nodes of the homogeneous  spin-1/2 chain governed by the XXZ
 Hamiltonian   in the presence of the  $z$-directed strong external inhomogeneous magnetic field  which is perpendicular to the chain:
\begin{eqnarray}\label{Hamiltonian}
&&
{\cal{H}}=\sum_{{i,j=1}\atop{j>i}}^{N}
D_{i,j}(I_{i,x}I_{j,x} + I_{i,y}I_{j,y}-2  I_{i,z}I_{j,z})+ \sum_{i=1}^N (\Omega_i+\Omega)I_{i,z} ,
\;\;\;D_{i,j}=\frac{\gamma^2 \hbar}{r_{i,j}^3},
\end{eqnarray}
where $\gamma$ is the gyromagnetic ratio,  $r_{i,j}$ is the distance between the $i$th and the $j$th spins,  $I_{i,\alpha}$ is the projection operator of the $i$th  spin on the $\alpha$ axis, $\alpha=x,y,z$,
$D_{i,j}$ is the dipole-dipole coupling constant between the $i$th and the $j$th nodes, 
$\Omega_i=\gamma h_i$, $\Omega=\gamma h$, $\vec h_i$ is the  inhomogeneous part of the magnetic field generated in the $i$th node by some special environment (for instance, by the system of direct currents, see Sec.\ref{Section:currents}), $\vec h$ is the external homogeneous magnetic field.  $h_i$ and $h>0$ are $z$-coordinates of $\vec h_i$ and $\vec h$ respectively. The field $\vec h$ must be strong enough in order to provide positivity of  the Larmor frequencies which are $\Omega_i+\Omega$, $i=1,\dots,N$.
This Hamiltonian describes  the secular part of the dipole-dipole interaction in the strong external  magnetic field
\cite{A}. Let $a$ be the distance between the neighboring nodes.
We denote $D_i\equiv D_{n,n+i}=\gamma^2\hbar/(i \, a)^3$, $n=1,2,\dots,N-i$, $i=1,2,\dots,N-1$. 

Hereafter, we will use the dimensionless quantities such as   time $\tau$,  
 coupling constants $d_{n}$, Larmor frequencies $\omega_n$  and distances $\xi_{n,m}$, which are defined as follows:
\begin{eqnarray}\label{tau}
&&
\tau= D_{1} t,
\;\;\;
d_{n}=\frac{D_{n}}{D_{1}}=\frac{1}{n^3},\;\;\;\omega_n=\frac{\Omega_n}{D_{1}},\;\;\;\omega=\frac{\Omega}{D_{1}},\;\;\;
\xi_{n,m}=\frac{r_{n,m}}{a}.
\end{eqnarray}
Using  definitions (\ref{tau}) and taking into account that $D_{1}=\gamma^2 \hbar/a^3$, the Hamiltonian (\ref{Hamiltonian}) may be written as follows:
\begin{eqnarray}\label{Hamiltonian_dl}
&&
{\cal{H}}=D_{1} \tilde {\cal{H}},\;\;\;
\tilde {\cal{H}}=\sum_{{i,j=1}\atop{j>i}}^{N}
d_{j-i}(I_{i,x}I_{j,x} + I_{i,y}I_{j,y}-2  I_{i,z}I_{j,z})+\sum_{i=1}^N (\omega_i+\omega) I_{i,z}.
\end{eqnarray}

It is important, that the Hamiltonian (\ref{Hamiltonian}) commutes with $I_z$ ($z$-projection of the total spin):
\begin{eqnarray}
[{\cal{H}},I_z]=0.
\end{eqnarray}
This means that both ${\cal{H}}$ and $I_z$ have the common set of eigenvectors. 
Since we  study the problem of single quantum state transfer, the  Hamiltonian is representable in the basis of $N$ eigenvectors
$|n\rangle$, $n=1,\dots,N$. Here Dirac notation $|j\rangle$ means that $j$th spin is excited, i.e. directed opposite to the external magnetic field. Then
the matrix representation $\tilde H$ of the Hamiltonian $\tilde{\cal{H}}$   is  following:
  \begin{eqnarray}\label{H1}
&&
\tilde H=\frac{1}{2}(D -\Gamma I),\\\nonumber
&&
D=\left(\begin{array}{ccccccc}
A_{1} & d_1 &d_{2} & \cdots& d_{N-3} & d_{N-2}&d_{N-1}\cr
d_1& A_{2} & d_1 & \cdots& d_{N-4} & d_{N-3}&d_{N-2}\cr
d_{2}&d_1& A_{3} & \cdots& d_{N-5} & d_{N-4}&D_{N-3}\cr
\vdots &\vdots &\vdots &\vdots&\vdots &\vdots &\vdots \cr
d_{N-3}&d_{N-4}&d_{N-5}&\cdots&A_{3}&d_{1}&d_{2}\cr
d_{N-2}&d_{N-3}&d_{N-4}&\cdots&d_{1}&A_{2}&d_1\cr
d_{N-1}&d_{N-2}&d_{N-3}&\cdots&d_{2}&d_1&A_{1}
\end{array}\right),
\\\nonumber
&&
A_{n}=2\left(\sum_{{i=1}\atop{i\neq n}}^{N}d_{|n-i|}+\omega_n+\omega\right),\;\;\Gamma=\sum_{{i,j=1}\atop{j> i}}^{N}d_{j-i},
\end{eqnarray}
where $I$ is the $N\times N$ identity matrix.

 We assume that  the $k_0$th node is excited initially.
In order to characterize the effectiveness of the quantum state transfer from the $k_0$th to the $n$th node, the fidelity $F_{k_0,n}(t)$ has been introduced  \cite{Bose}:
\begin{eqnarray}\label{fidelity}
F_{k_0,n}(\tau)=\frac{|f_{k_0,n}(\tau)|\cos\Gamma_{k_0,n}(\tau)}{3}+
\frac{|f_{k_0,n}(\tau)|^2}{6}+\frac{1}{2},\;\;\Gamma_{k_0,n}=\arg f_{k_0,n}
,
\end{eqnarray}
where $f_{n,m}$ is the transition amplitude:
\begin{eqnarray}\label{fnm}
 f_{n,m}&=&\langle m |e^{-i\tilde {\cal{H}} \tau}|n\rangle=\sum_{j=1}^{N}  u_{n,j} u_{m,j}e^{-i\lambda_j  \tau/2},\;\;f_{n,m}=f_{m,n}.
 \end{eqnarray}
 Here $u_{i,j}$, $i,j,=1,\dots,N$, are components of the normalized eigenvector
$u_j$ corresponding to the  eigenvalue $\lambda_j$ of the matrix
$D$:
$Du_j=\lambda_j u_j$.
It was noted  \cite{Bose} that any particular $\Gamma_{i,j}$ 
 may be put to zero by the proper choice of the constant homogeneous
 magnetic field (if the end-to-end state transfer is considered 
 then  $\Gamma_{1N}$ may be put to zero). This remark has been generalized in 
 \cite{FZ}, where all $\Gamma_{i,j}$  have been compensated by 
 the proper choice of the time-dependent homogeneous magnetic field, i.e. $\omega$ must be replaced by $\tilde \omega(\tau)$ in eq.(\ref{Hamiltonian_dl}). However we do not consider the problem of compensation of $\Gamma_{i,j}$ (a solution of this problem is given in \cite{FZ}) and take  $\omega$ independent on $\tau$  as well as  $\Gamma_{i,j}=0$ in eq.(\ref{fidelity}). Then the fidelity $F_{k_0,n}$ is uniquely defined by  $|f_{k_0,n}|$. For this
 reason, we
  consider the probability  $P_{k_0,n}=|f_{k_0,n}|^2$ instead of 
  $F_{k_0,n}$ as the characteristic of the quantum state transfer 
  from the $k_0$th to the $n$th node. The fact that we are going to consider the exited state transfer (instead of arbitrary state transfer) justifies our choice of the basis consisting of $N$ eigenvectors. Otherwise this basis would consist of $N+1$ vectors  including the eigenvector $|0\rangle$ corresponding to the  ground state.
In addition, as far as the homogeneous part of the magnetic field does not effect the probability of the state transfer,   we take $\omega=0$ for the sake of simplicity and refer to $\omega_i$ as Larmor frequencies.

It is known \cite{KS} that the PST  requires, for some time $\tau_0$,
 \begin{eqnarray}\label{perfect}
 e^{-i\tilde {\cal{H}} \tau_0} |n\rangle = e^{i \phi_0}|m\rangle \;\;\Rightarrow \;\;
 \lambda_k \tau_0  = (2  n_k + k) \pi + \phi_0,
\end{eqnarray} 
where $\phi_0$ is some arbitrary phase and $n_k$ are arbitrary integers. This  is a simple condition for eigenvalues $\lambda_k$, 
but it is hard to reconstruct the full original Hamiltonian $\tilde{\cal{H}}$ corresponding to these eigenvalues \cite{Kay}. This problem may be correctly solved only for the approximation of nearest neighbor interaction \cite{KS}, which is not considered here. 
Regarding the analytical study of the HPSTs, the situation is even more complicated since condition (\ref{perfect}) is not satisfied precisely in this case. 
 For this reason we use numerical simulations to find Larmor frequencies $\omega_i$  which provide good parameters of HPSTs.

We will use the Wootters criterion with the concurrence $C_{n,m}$ as the measure of entanglement between the $n$th and the $m$th nodes. In this case,  the following formula is valid:
\begin{eqnarray}
C_{n,m}(\tau)=2 \sqrt{P_{k_0,n}(\tau) P_{k_0,m}(\tau)},
\end{eqnarray}
relating the concurrence with the excited state transfer probabilities \cite{GMT,AOPFP,DFZ}.
Here  $n$, $m$ and $k_0$ are arbitrary nodes of the chain.

The following general notations will be used throughout this paper:
\begin{itemize}
\item
  $\bar P_{k_0,m}\equiv P_{k_0,m}( \tau^P_{k_0,m}) $ and  $\tau^P_{k_0,m}$ for the  probability of the HPST  and  the time interval required for the  HPST  from the $k_0$th to the $m$th node of the $N$-node chain 
\item
  $\bar C_{k_0,m}\equiv C_{k_0,m}( \tau^C_{k_0,m})$ and  $\tau^C_{k_0,m}$ for the  concurrence and   the time moment when the $k_0$th and $m$th nodes  become strongly entangled.
\item
  The HPST from the $k_0$th  to the $n$tn node is referred to as HPST$_{k_0,n}$.
\item
We refer to the set of parameters
\begin{eqnarray}
\omega_i,\;\;i=1,\dots,N,\;\; \bar P_{n,m},\;\;\tau^P_{n,m},\;\; \bar C_{n,m},\;\;\tau^C_{n,m}
\end{eqnarray}
as the parameters of the HPST$_{n,m}$.
\end{itemize}

We will see that Larmor frequencies   $\omega_i$, $i=1,\dots,N$, providing HPST$_{n,m}$ with short enough $ \tau^P_{n,m}$ do not  guaranty the strong entanglement between the $n$th and the $m$th during the time interval comparable with $ \tau^P_{n,m}$, i.e. it is possible that $\tau^C_{n,m} \gg \tau^P_{n,m}$.  We call such transfers as HPST$^P_{n,m}$. We show that, taking other values of $\omega_i$, one can generate HPST$_{n,m}$ such that  the $n$th and the $m$th nodes become strongly entangled at the moment  $\tau^C_{n,m}$ and  $\tau^P_{n,m} \gg \tau^C_{n,m}$. These transfers will be called HPST$^C_{n,m}$. The HPST$^{P}_{n,m}$ which provide the strong entanglement between the $n$th and the $m$th nodes with $\tau^P_{n,m}\approx 2 \tau^C_{n,m}$  will be referred to as HPST$^{PC}_{n,m}$.  

\subsection{Inhomogeneous magnetic field}
\label{Section:mf}
\label{Section:currents}
We describe a variant of the practical realization of the proper inhomogeneous 
magnetic field along the spin chain in this section.
Let $\vec j_1$ be the direct current directed along $x$-axis which is  perpendicular to both the spin chain and the $z$-axis (so that the  magnetic field of this current is  $z$-directed). 
The $z$-coordinate   $h_1$ of the magnetic field generated by this current is following:
\begin{eqnarray}
h_1 =\frac{B_1 }{ r_1},\;\;B_1=\frac{2 j_1}{c},
\end{eqnarray} 
where  $j_1$ is the $x$-coordinate of $\vec j_1$  (which may be either positive or negative)
$c$ is the light velocity 
$ r_1=|\vec r_1|$ is the distance between the  given point and the  wire with current. 

\begin{figure*}
   \epsfig{file=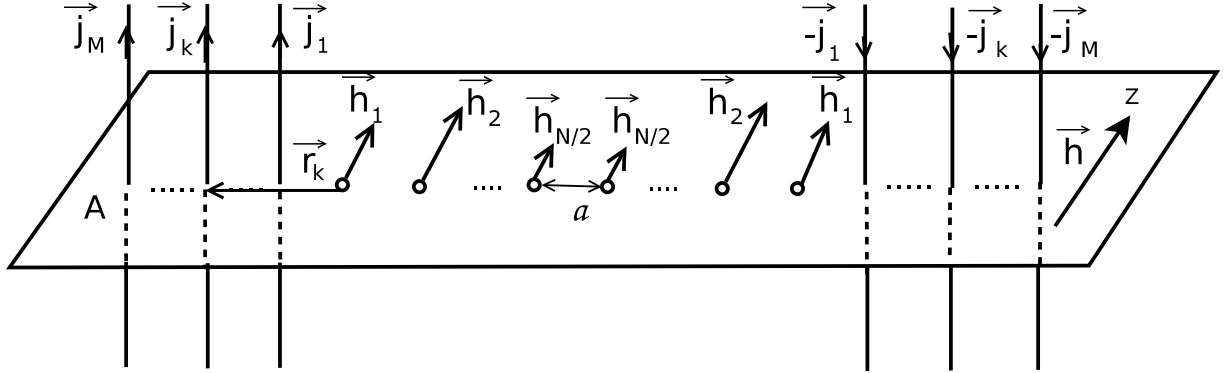
   , scale=0.5,angle=0}
  \caption{The system of direct currents $\vec j_1,\dots, \vec j_M$ and $-\vec j_1,\dots, -\vec j_M$ generating a symmetrical distribution of the  magnetic fields  $\vec h_j$, $j=1,\dots,N/2$ in the nodes of the $N$-node spin chain.  $N$ is even in this figure, $\vec h$ is an external homogeneous magnetic field.
  \label{Fig:currents} }
\end{figure*}

Of course, using the system of direct currents, we are able to generate more complicated magnetic field. 
For instance, let our chain belong to  some plane $A$ and all currents are directed  perpendicular to this plane, see Fig.\ref{Fig:currents}.
It is simple to see that  one can  generate arbitrary $\Omega_m$, $m=1,\dots,N$, in all nodes of the spin chain using the system of $N$ currents. 
In order  to provide the symmetry
\begin{eqnarray}
h_n=h_{N-n+1},\;\;n=1,\dots, N_0,\;\;N_0=\left\{\begin{array}{ll}
N/2,\;\;{\mbox{even}}\;\; $N$\cr
(N+1)/2,\;\;{\mbox{odd}}\;\; $N$
\end{array}\right.
,
\end{eqnarray} 
which is required for the end-to-end state transfer,
we take the $M=N_0$ currents in one side of the chain and $N_0$ symmetrically placed but oppositely  directed  currents  in another side of the chain.  All currents cross the plane $A$ along the line containing the chain. Thus 
\begin{eqnarray}\label{Omega}
h_m &=& \sum_{k=1}^{N_0} \left(\frac{B_k}{r_k -(m-1)a} +\frac{B_k}{r_k +(N-m)a}\right),\;\;\Rightarrow\\\label{Omega2}
\omega_m&=&\sum_{k=1}^{N_0} \left(\frac{b_k}{\xi_k -(m-1)} +\frac{b_k}{\xi_k +(N-m)}\right),\\\nonumber
&&
b_k=\frac{B_ka^2}{\gamma \hbar  },\;\;\xi_k=\frac{r_n}{a}
,\;\;B_k=\frac{2 j_k}{c},\;\;m=1,\dots,N_0
,
\end{eqnarray}
where $j_k=|\vec j_k|$,
$r_k\equiv |\vec r_k|$ is the distance between the  current $\vec j_k$  and the first node, $k=1,\dots,N_0$, see Fig.\ref{Fig:currents}.

\section{HPSTs and entanglements  between different nodes of the spin chains}
\label{Section:examples}

We describe   HPSTs and entanglements between different  nodes in chains of different length.
The simplest nontrivial  chain is the three-node one. Parameters of different HPSTs are given in Table I.
\begin{table*}[!htb]
\begin{tabular}{|p{0.5cm}|p{2.cm}|p{1.5cm}|p{1.5cm} |p{1.5cm}|p{1.2cm}|p{1.2cm}|p{1.2cm}|p{1.2cm}|}
\hline
n.& $HPST$s                 &$\omega_1$&$\omega_2$&$\omega_3$&$\bar P$ & $ t^P$ & $\bar C$ & $ t^C$ \\\hline
1&$HPST^{PC}_{1,2}$&15.891   &15.000        &0         &0.999    &3.126       &1.000     &1.564\\\hline
2&$HPST^{P}_{1,3}$ &1.063    &0         &1.063     &1.000    &4.375       & ---      & $>4400$\\\hline
3&$HPST^{PC}_{1,3}$ &1.979    &0         &1.979     &1.000    &6.855       &1.000     &3.428\\\hline
\end{tabular}
\label{Table:HPST3}
\caption{Parameters of the  HPSTs in the  three-node spin-1/2 chain}
\end{table*}
Parameters of  HPSTs  in the four-node chain are represented in Table II.
\begin{table*}[!htb]
\begin{tabular}{|p{0.5cm}|p{2.cm}|p{1.5cm}|p{1.5cm} |p{1.5cm}|p{1.2cm}|p{1.2cm}|p{1.2cm}|p{1.2cm}|p{1.2cm}|}
\hline
n.& $HPST$s                &$\omega_1$&$\omega_2$&$\omega_3$&$\omega_4$&$\bar P$ & $ t^P$ & $\bar C$ & $ t^C$ \\\hline
1&$HPST^{PC}_{1,2}$&50.968   &50.000        &0         &0         &1.000    &3.136       &1.000     &1.570\\\hline
2&$HPST^{PC}_{1,3}$ &2.107    &0         &1.028     &0&         0.943    &6.798       & 0.964    & 3.411  \\\hline
3&$HPST^{P}_{1,4}$ &1.171    &0         &0         &1.171&    0.982    &5.516       &---     &$>30$\\\hline
4&$HPST^{PC}_{1,4}$ &2.465    &0         &0         &2.465&    0.961    &16.822      &0.990   &8.427\\\hline
5&$HPST^{PC}_{2,3}$ &0        &175.000         &175.000         &0&    1.000   &3.145       &1.000   &1.572\\\hline
\end{tabular}
\label{Table:HPST4}
\caption{Parameters of the HPSTs in the four-node spin-1/2 chain}
\end{table*}
 The three- and four-node chains represent examples of simplest spin systems allowing one to destribute the quantum state  between several receivers as well as to organize strong entanglement between any two nodes, which is shown in 
 Tables I and II.

\begin{figure*}
   \epsfig{file=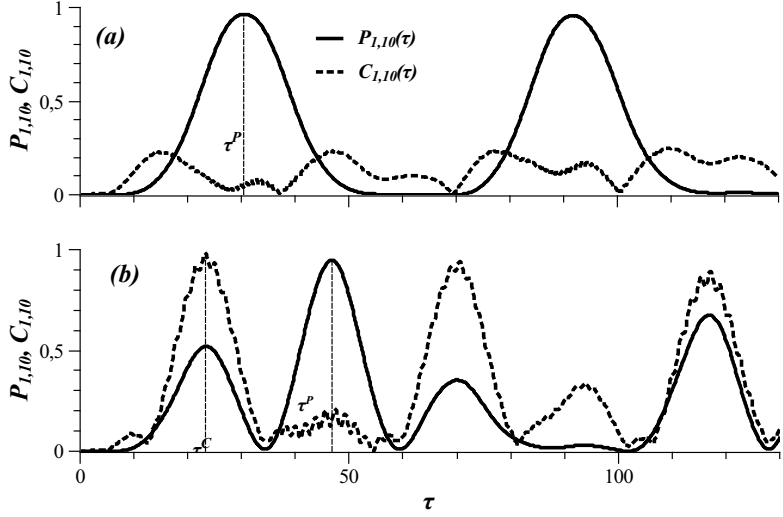 
   , scale=0.6,angle=270}
  \caption{The probability of the end-to-end state transfer and the end-to-end concurrence in the ten-node spin chain with nonzero $\omega_1=\omega_{10}$, $\omega_2=\omega_{9}$ and $\omega_3=\omega_{8}$, 
$\tau^P\equiv \tau^P_{1,10}$, $\tau^C\equiv \tau^C_{1,10}$; ($a$) 
HPST$^P_{1,10}$: $\omega_1=2.185$, $\omega_2=-5.585$, $\omega_3=0.688$, $\bar P_{1,10}=0.965$, $\tau^C>1600$, $\tau^P=30.543$; ($b$) 
 HPST$^{PC}_{1,10}$:  $\omega_1=2.314$, $\omega_2=-1.816$, $\omega_3=0.916$, $\bar P_{1,10}=0.949$, $\bar C_{1,10}=0.984$, $\tau^P=46.728$, $\tau^C=23.263$.
  \label{Fig:N10_3} }
\end{figure*}

Similarly, one can consider state transfers and entanglements between any two nodes in longer chains. However, 
we concentrate on the end-to-end state transfer and entanglement. 
We consider four  examples of the HPSTs between the end nodes of the ten-node chain, 
see Table III. The first example corresponds to 
all zero $\omega_i$, exept $\omega_1=\omega_{10}$, which is single optimization parameter
 (see the first row in the table); the second example corresponds to all zero $\omega_i$ except 
$\omega_1=\omega_{10}$, $\omega_2=\omega_{9}$, which are two optimization parameters 
(see the second and the third rows in the table); the third example corresponds to all zero $\omega_i$ except
$\omega_1=\omega_{10}$, $\omega_2=\omega_{9}$, $\omega_3=\omega_{8}$, which are three optimization parameters 
(see the 4th and the 5th rows in the table). 
Four last columns of Table III show us that each subsequent example  improves parameters of the HPST. One must expect that the best 
parameters will correspond to all nonzero $\omega_i$ having proper values. However, this is very complicated optimization problem which is not considered here.
\begin{table*}[!htb]
\begin{tabular}{|p{0.5cm}|p{2.cm}|p{1.5cm}|p{1.5cm} |p{1.5cm}|p{1.2cm}|p{1.5cm}|p{1.2cm}|p{1.5cm}|}
\hline
n.& $HPST$s                 &$\omega_1=\omega_{10}$&$\omega_2=\omega_9$&$\omega_3=\omega_8$&$\bar P_{1,10}$ & $ t^P_{1,10}$ & $\bar C_{1,10}$ & $ t^C_{1,10}$ \\\hline
1&$HPST^{PC}_{1,10}$&2.651   &0         &0                 &0.971    &330.352 &0.944     &165.275\\\hline
2&$HPST^{P}_{1,10}$ &2.133    &-12.435   &0        &         0.992    &59.776      & ---      &$>7500$ \\\hline
3&$HPST^{PC}_{1,10}$ &2.192    &-10.435   &0         &    0.988    &104.271     &0.996   &51.863\\\hline
4&$HPST^{P}_{1,10}$ &2.185    &-5.585    &0.688     &    0.965    &30.543      & ---    &$>1600$\\\hline
5&$HPST^{PC}_{1,10}$ &2.314    &-1.816      &0.916      &    0.949   &46.728      &0.984   &23.263\\\hline
6&$HPST^{P}_{1,10}$ &\multicolumn{3}{p{3.6cm}|}{mag. field of two currents, $b=1011.150$}&    0.972    &171.045     & ---    &$>8200$\\\hline
7&$HPST^{C}_{1,10}$ & \multicolumn{3}{p{3.6cm}|}{ $b=1527.500$} &    ---     &$>1300$     &0.943   & 180.912\\\hline
\end{tabular}
\label{Table:HPST10}
\caption{Parameters of the end-to-end  HPSTs in the ten-node spin-1/2 chain}
\end{table*}
To demonstrate general behavior of probabilities and concurrences we show proper curves 
 for the case of three pairs of nonzero Larmor frequencies in Fig.\ref{Fig:N10_3}.

   We have all fixed Larmor frequencies which are either zero or nonzero in three above examples. 
This may be done in the experiment using a system of ten currents (five currents in each side of the chain, see Fig.\ref{Fig:currents} where $M=5$, $N=10$) and solving the system (\ref{Omega2}) with $N_0=5$ for $b_i$, $i=1,\dots,5$. 
 The next example 
demonstrates 
a simple method allowing one to  organize HPST$^P_{1,10}$ and HPST$^C_{1,10}$ in the magnetic field of two direct currents symmetrically  placed with respect to the spin chain, i.e. $M=1$ in  Fig.\ref{Fig:currents} and $N_0=1$ in eq.(\ref{Omega2}).
In accordance to  eq.(\ref{Omega2}),
the  distribution of $\omega_n$ generated by  one pair of the direct currents is following ($b_1\equiv b$):
\begin{eqnarray}
\omega_n=\omega_{10-n+1},\;\;\omega_n=b\left(
\frac{1}{\xi+n-1}+\frac{1}{\xi+10-n}
\right),\;\;n=1,\dots,5.
\end{eqnarray}
The parameters of HPSTs corresponding to  fixed $\xi=20$ and optimized  $b$ 
are represented in Table III, see the 6th and the 7th rows.
 We see that  the strong end-to-end entanglement does not assume the end-to-end HPST during reasonable time interval and vise-verse. Parameters $\tau^P_{1,10}$,  $\tau^C_{1,10}$ are better while parameters $\bar P_{1,10}$ and $\bar C_{1,10}$ are comparable with  those which have been found in  the first example (see the first row in  Table III).

Similarly, we represent the results of the same four experiments in the twenty-node chain, see Table IV. 
\begin{table*}[!htb]
\begin{tabular}{|p{0.5cm}|p{2.cm}|p{1.5cm}|p{1.5cm}|p{1.5cm}|p{1.2cm}|p{1.7cm}|p{1.2cm}|p{1.7cm}|}
\hline
n.& $HPST$s               & $\omega_1=\omega_{20}$&$\omega_2=\omega_{19}$&$\omega_3=\omega_{18}$&$\bar P_{1,20}$ & $t^P_{1,20}$ & $\bar C_{1,20}$ & $ t^C_{1,20}$ \\\hline
1&$HPST^{PC}_{1,20}$&3.615   &0         &0                 &0.975    &10209.184&0.961     &5184.003\\\hline
2&$HPST^{PC}_{1,20}$ &1.569    &-67.000   &0        &         0.989    &160.615     &0.974     &79.857 \\\hline
3&$HPST^{P}_{1,20}$ &1.567    &-6.300   &0.255     &    0.949    &62.377      & ---    &$>1400$\\\hline
4&$HPST^{C}_{1,20}$ &1.742    &-4.200    &0.549     &    ---      &$>5300$     &0.923   &47.126 \\\hline
5&$HPST^{PC}_{1,20}$ & \multicolumn{3}{p{3.6cm}|}{mag. field of two currents, $b=1367.000$}&
                                                       0.975   &10158.681   &0.994   &5119.620\\\hline
\end{tabular}
\label{Table:HPST20}
\caption{Parameters of the end-to-end  HPSTs in the twenty-node spin-1/2 chain}
\end{table*}
We see that the time intervals $t^P$ and $t^C$ corresponding to the 10-node chains are about 30 times shorter then 
those corresponding to the 20-node chain in the case of single nonzero pair of Larmor frequencies (compare the first lines of Tables III and IV), while these parameters defers by the factor about 2 in the case of three nonzero pairs of Larmor frequencies  (compare the 4th and the 5th rows of Table III with the 3rd and the 4th rows of   Table IV).
Note also that 
 parameters of the HPSTs found in the  experiment with the magnetic field generated by the system of two currents  
   are comparable with  those which have been found in the  experiment with single pair of nonzero Larmor frequencies 
(compare  the fifth and the first rows in  Table IV).

\subsection{On the optimization problem}
As we have seen, using the specific distribution of the external 
inhomogeneous magnetic field (i.e. the specific values of the Larmor frequencies $\omega_i$) one can 
provide the HPSTs and/or strong entanglements between two different nodes of  spin chains.
 These specific values $\omega_i$ may be found as a result of optimization problems. 
Doing such optimization, we have, first of all, to fix a time interval $T$
 during which we want to detect a signal 
(i.e. to observe the high amplitude of either  $P(t)$ or  $C(t)$). The value of $T$ depends on the length of the chain. Thus,
in the case shown in Fig.\ref{Fig:N10_3} one can take $T=50$. In general, if $S$ is some set of nodes in spin system  and we want to 
 obtain HPST$^P$ and /or HPST$^C$ between any two nodes of $S$,  them  $T\ge \max\limits_{n,m\in S}(t^P_{nm},t^{C}_{nm})$.

Most simple is optimization 
over single parameter, see  the second and the third rows in Table I,  
the  3rd-5th rows in Table II, the first, the 6th and the 7th rows in Table III,   the first and 
the 5th rows in Table IV.
Starting with zero value of the optimization parameter and increasing it by step $0.001$, we have found such value, which yields satisfactory values of desiable HPSTs parameters, i.e. $\bar P_{n,m}$ and $t^{P}_{n,m}$ 
(or $\bar C_{n,m}$ and $t^{C}_{n,m}$) for some  $n$ and $m$. 

Optimization over two parameters  is more tricky 
(see  the first row in Table I, the first and  the second rows in Table II, 
the   second  and the third rows in Table III, the second row in Table IV). 
One has to scan a two-dimensional space of optimization parameters in this case. 

Optimization over three parameters is mostly complicated process used in this paper, 
see  the   4th and the 5th rows in Table III, the 3rd and the 4th rows in Table IV. One has to scan a tree dimensional space of optimization parameters.

Doing each optimization process one has to compromise between the  amplitudes $\bar P$, $\bar C$ 
(which must be as high as possible) and the time intervals $t^P$, $t^C$  (which must be as short as possible). Parameters represented in Tables I-IV  are results of appropriate optimization and "compromise" problems. In other  words, our optimization problems do not have unique solutions, so that  there are other sets of parameters of HPSTs which  defer from those represented in the above Tables. We have found only particular solutions to the problem of the state transfer and entanglemet organization. 

Finally note that the feasible length of the spin chains is much bigger than twenty nodes. It is at least 
 several hundreds nodes. Numerical study of such long chains with dipole-dipole interactions between all nodes is possible due to the fact that we consider the single state transfer in the system governed by the Hamiltonian commuting with the total spin projection $I_z$. This reduces the dimensionality of the Hamiltonian from $2^N$ to $N$ ($N$ is the number of nodes).  However, the optimization of long chains is time consuming process which is not considered in this paper.

\section{Conclusions}
\label{Section:conclusions}

We demonstrate  that the   external inhomogeneous magnetic field may be effectively used in order to arrange the both HPSTs and entanglements between any two  nodes in  spin-1/2 chains. This is an alternative method to the traditional one where the end-to-end HPST is possible due to the proper values of the  coupling constants. The basic advantage of our method is that we minimize requirements to the spin chain as the basic object of a communication channel  because  a homogeneous chain is most simple for the realization in practice. However, one has to  prepare the proper distribution of the magnetic field using some additional environment, which must be done for each particular length of the chain. We suggest a method to create such magnetic field using a system of direct currents properly placed with respect to the chain. Although we pay most attention to the end-to-end state transfer and entanglement, both the
 HPST and the entanglement between any two nodes may be organized, which is shown in simple examples of the three- and the four-node chains.

The summary of basic results is below.
\begin{enumerate}
\item
We suggest a method organizing the quantum state transfer between different nodes of spin-1/2 chains 
 based on the homogeneous chains governed  by the XXZ Hamiltonian  in the presense of  the proper inhomogeneous magnetic field. This is an alternative method to that based on the inhomogeneous chains with homogeneous magnetic field \cite{CDEL,ACDE,KS,GKMT}. 
\item
The proper distribution of the magnetic field may be achieved using the system of direct currents, see Sec.\ref{Section:mf}.
\item
Using the inhomogeneous magnetic field, we arrange the HPST$^P$ between any two nodes of the spin chain (not only between the end nodes, see \cite{CDEL,ACDE,KS,GKMT}). Since this question is not trivial even for short chains, we consider the state transfer between all nodes in the chains of three and four nodes, Tables I and  II. 
Regarding the chains of ten and twenty nodes, we obtain only parameters of the end-to-end HPSTs, 
 Tables III and  IV
\item
The inhomogeneous magnetic field allows one to realize  the
strong entanglement between  the $n$th and $m$th  nodes of the chain. 
 We arrange the HPST$^C$ between any two nodes of the three- and four-node chains (see Tables I and  II) and between the end nodes of the ten- and twenty-node chains 
 (see  Tables III and  IV).
\item
Considering the end-to-end state transfer and entanglement in the ten- and twenty-node chains we observe that parameters of such transfers may be improved involving more and more pairs of nonzero Larmor frequencies $\omega_i$,
 see 1st-5th rows of Table III and 1st-4th rows of Table IV.
\item
We use interactions between all nodes of the chain throughout this paper, 
which is important for the chains with dipole-dipole interactions. For instance, one can demonstrate that the approximation of nearest neighbor interactions 
 produces much much longer state transfer time intervals, i.e. does not yield the correct results. 
\end{enumerate}

Authors thank Professor E.B.Fel'dman for useful discussions.
This work is supported by the Russian Foundation for Basic Research through the grant 07-07-00048 and by the Program of the Presidium of RAS No.18.

\end{document}